\documentstyle[a4p,12pt,epsfig,cite,rotating]{article}

\begin{document}
%========================================================================
%
%  define a few new names (remember math mode ne text mode!)
%
\newcommand{\xE}{\mbox{$x_E$}}
\newcommand{\xp}{\mbox{$x_p$}}
\newcommand{\lnxp}{\mbox{$\log(1/x_p)$}}
\newcommand{\degree}{$^{\circ}\space$}
%
%--    ... Particles
%
\newcommand{\epem}{\mbox{$\mathrm{e^+e^-}$}}
\newcommand{\Zzero}{\mbox{${\mathrm{Z}^0}$}}
\newcommand{\azpm}{\mbox{$\mathrm{a}_0^{\pm}$}}
\newcommand{\rpm}{\mbox{$\rho^{\pm}$}}
\newcommand{\rpml}{\mbox{$\rho(770)^{\pm}$}}
\newcommand{\piz}{\mbox{${\pi^0}$}}
\newcommand{\pipm}{\mbox{$\pi^{\pm}$}}
%
%--   ... Decay channels
%
\newcommand {\threepi} {\mbox{$\pi^+\pi^-\pi^0$}}
\newcommand {\rhopipi} {$\rho^{\pm} \rightarrow \pi^{\pm}\pi^0$}
\newcommand {\omegapipipi} {$\omega \rightarrow \pi^+\pi^-\pi^0$}
\newcommand {\omreslall} {0.312 $\pm$ 0.011 $\pm$ 0.030}
\newcommand {\omresall}  {0.312 $\pm$ 0.032}
\newcommand {\omreslhi} {0.142 $\pm$ 0.081 $\pm$ 0.080}
\newcommand {\omreshi}  {0.142~$\pm$~0.114}
\newcommand {\rhoreslall} {0.245 $\pm$ 0.035 $\pm$ 0.090}
\newcommand {\rhoresall}  {0.245 $\pm$ 0.097}
\newcommand {\rhoreslhi} {0.373 $\pm$ 0.035 $\pm$ 0.038}
\newcommand {\rhoreshi}  {0.373 $\pm$ 0.052}
%
%  .025 - 0.600 &  .332 &  .049 &  .022 &  .017 &  .010 &  .009 &  .026 &  .063 \\
%  .025 - 0.300 &  .322 &  .051 &  .020 &  .021 &  .010 &  .009 &  .027 &  .066 \\
%
%----------------------
 
%----- Title page -----
%----------------------
\begin{titlepage}

\begin{center}
{\large   EUROPEAN LABORATORY FOR PARTICLE PHYSICS }
\end{center}
\bigskip
\begin{flushright}
       CERN-EP/99-082 \\
       14th June 1999
\end{flushright}

\vfill

\begin{center}
{\LARGE\bf  A Study of Spin Alignment of
            {\boldmath $\rho(770)^{\pm}$} and
            {\boldmath $\omega(782)$} Mesons
            in Hadronic Z{\boldmath$^0$} Decays }
\end{center}

\vfill
\begin{center}{\LARGE The OPAL Collaboration}\end{center}
\vfill

\begin{center}{\Large\bf Abstract}
\end{center}

The helicity density matrix elements $\rho_{00}$ of \rpml\
and $\omega(782)$ mesons
produced in Z$^0$ decays have been measured using the
OPAL detector at LEP.
Over the measured meson energy range,
the values are compatible with 1/3,
corresponding to a statistical mix of helicity $-1$, 0 and $+1$ states.
For the highest accessible scaled energy range 0.3 $<$ $x_E$ $<$ 0.6,
the measured $\rho_{00}$ values of the \rpm\
and the $\omega$ are
\rhoreshi\   and
 \omreshi, respectively.
These results are compared to measurements of other
vector mesons.

\vfill

\begin{center}
{\large (To be submitted to European Physical Journal C) }
\end{center}

\vfill

\end{titlepage}

%========================================================================

\newpage

\begin{center}{\Large        The OPAL Collaboration
}\end{center}\bigskip
\begin{center}{
%begin authorlist PLEASE DO NOT DELETE THIS COMMENT
G.\thinspace Abbiendi$^{  2}$,
K.\thinspace Ackerstaff$^{  8}$,
G.\thinspace Alexander$^{ 23}$,
J.\thinspace Allison$^{ 16}$,
N.\thinspace Altekamp$^{  5}$,
K.J.\thinspace Anderson$^{  9}$,
S.\thinspace Anderson$^{ 12}$,
S.\thinspace Arcelli$^{ 17}$,
S.\thinspace Asai$^{ 24}$,
S.F.\thinspace Ashby$^{  1}$
% ...
%end authorlist PLEASE DO NOT DELETE THIS COMMENT
}\end{center}\bigskip
\bigskip
%begin institutes
$^{  1}$School of Physics and Astronomy, University of Birmingham,
Birmingham B15 2TT, UK
\newline
$^{  2}$Dipartimento di Fisica dell' Universit\`a di Bologna and INFN,
I-40126 Bologna, Italy
\newline
% ...
$^{  5}$Cavendish Laboratory, Cambridge CB3 0HE, UK
\newline
% ...
$^{  8}$CERN, European Organisation for Particle Physics,
CH-1211 Geneva 23, Switzerland
\newline
$^{  9}$Enrico Fermi Institute and Department of Physics,
University of Chicago, Chicago IL 60637, USA
\newline
% ...
$^{ 12}$Indiana University, Department of Physics,
Swain Hall West 117, Bloomington IN 47405, USA
\newline
% ...
$^{ 16}$Department of Physics, Schuster Laboratory, The University,
Manchester M13 9PL, UK
\newline
$^{ 17}$Department of Physics, University of Maryland,
College Park, MD 20742, USA
\newline
% ...
$^{ 23}$Department of Physics and Astronomy, Tel Aviv University,
Tel Aviv 69978, Israel
\newline
$^{ 24}$International Centre for Elementary Particle Physics and
Department of Physics, University of Tokyo, Tokyo 113-0033, and
Kobe University, Kobe 657-8501, Japan
\newline
% ...
$^{  a}$ and at TRIUMF, Vancouver, Canada V6T 2A3
\newline
% ...
$^{  e}$ and University of Mining and Metallurgy, Cracow
\newline
% ...
\newline
%end notes

%========================================================================

%-------------------------
%----- Document text -----
%-------------------------

\newpage

%%%%%%%%%%%%%%%%%%%%%%%%%%%%%%%%%%%%%%%%%%%%%%%%%%%%%%%%%%%%%%%%%%%%%%%%%%%%

\section{Introduction}
\label{sect-intro}

Very little is known about the role of spin in the hadronization process.
At LEP, this can be investigated by studying the properties of vector
mesons produced in hadronic \Zzero\  decays.
Recent data on the helicity states of vector mesons
produced in hadronic \Zzero\
decays~\cite{bib-aliopal,bib-alidelphi,bib-aliopks,bib-alibstar}
reveal that the spin of high-energy K$^*(892)^0$ and
$\phi(1020)$ mesons is preferentially aligned transverse to the
direction of their momentum.
Measurements of the $\rho(770)^0$ and D$^{*}(2010)^{\pm}$ mesons
are consistent with this behaviour, while B$^*$ mesons
show no alignment.
The mechanism at the origin of spin alignment is not well
understood theoretically~\cite{bib-theo}, and this phenomenon is
ignored in models such as the Lund string model~\cite{bib-jetset}
and the cluster model~\cite{bib-herwig}.
Extending these studies to other vector mesons would provide
valuable help in elucidating the role of spin in hadronization.
The helicity density matrix elements for the \rpml\
and $\omega(782)$ mesons produced in \Zzero\  decays have
never been measured: these are particularly interesting
because the most pronounced alignments observed so far
are for light mesons.
The \rpm\  and $\omega$ mesons,
together with the $\rho^0$ whose helicity matrix
element has already been measured~\cite{bib-alidelphi},
have a similar quark structure and can be expected to
have similar behaviour.
Experimentally, the systematic uncertainties affecting the
extraction of the helicity matrix elements for these three
mesons differ substantially.

This paper describes the measurement of the
helicity density matrix elements $\rho_{00}$ of $\rho^{\pm}$
and $\omega$ mesons produced in Z$^0$ decays using the
OPAL detector at LEP.
The detector and its performance are described in detail
in Refs.~\cite{bib-opaldet,bib-opalsi,bib-dedx}.
The data sample consists of 4.1 million hadronic \Zzero\
decays collected at centre-of-mass energies
within $\pm$2 GeV of the \Zzero\  peak.
The selection of hadronic events is presented in
Ref.~\cite{bib-opalline}.
The method to reconstruct and identify the \rhopipi\  and
\omegapipipi\  decays is the same as used in
Ref.~\cite{bib-opalpiz}, where it is explained in detail.
The measurements of $\rho_{00}$ for the \rpm\
and $\omega$ mesons are presented in Sections 2 and 3,
and the results are compared to those for other mesons at LEP
in Section 4.

%%%%%%%%%%%%%%%%%%%%%%%%%%%%%%%%%%%%%%%%%%%%%%%%%%%%%%%%%%%%%%%%%%%%%%%%%%%%
                                                           
\section{ Helicity density matrix element 
          {\boldmath $\rho_{00}$} 
          of {\boldmath $\rho^{\pm}$} mesons }
\label{sect-anarho}

The \rpm\  resonance decays dominantly via the $\pi^{\pm}\pi^0$ channel.
Using as a spin analyser the angle in the $\pi^{\pm}\pi^0$
rest frame between one of the pion momenta and
the \rpm\  boost direction, the distribution of this angle,
$\theta_H$, is~\cite{bib-coseqn}:
\begin{eqnarray}
\label{eqn-coseqn}
W(\cos\theta_H) & = & \frac{3}{4}
         [ (1-\rho_{00}) + (3\rho_{00}-1)\cos^2\theta_H ]
\end{eqnarray}
where $\rho_{00}$ is the helicity density matrix element
expressing the probability that the spin of the \rpm\
meson be perpendicular to its momentum direction.
%Because of unitarity and parity conservation
%in the decay $\rho^{\pm}\rightarrow\pi^{\pm}\pi^0$,
%the probabilities that the spin
%be parallel or anti-parallel, $\rho_{11}$ and $\rho_{-1-1}$, are
%both equal to $(1-\rho_{00})/2$.

As the mechanisms producing the observed mesons
and their amount of spin alignment
can vary with their energy,
the analysis is repeated for different intervals of the
scaled energy $x_E=E_{\mbox{\scriptsize{meson}}}/
                   E_{\mbox{\scriptsize{beam}}}$.

\subsection{{\boldmath $\rho^{\pm}$} meson reconstruction}

The reconstruction of the decay \rhopipi\
follows exactly that described in Ref.~\cite{bib-opalpiz}.
Charged pion candidates are selected as tracks in the central
drift chambers with an energy loss measurement having a
probability greater than 1\% for the pion
hypothesis~\cite{bib-dedx}.
Neutral pions are obtained from the combination of
pairs of photons detected  either as localised
energy deposits in the electromagnetic calorimeter or as two tracks
from a $\gamma\rightarrow$ \epem\  conversion within the volume of the
central drift chambers,
and selected using the multivariate
method described in Ref.~\cite{bib-opalpiz}.

All selected \piz\ and $\pi^{\pm}$ candidates are combined
in pairs.
From the energy and momenta of the \piz\  and $\pi^{\pm}$ in each pair,
three quantities are evaluated:
the scaled energy of the pair,
\xE, the invariant mass of the $\pi^{\pm}\pi^0$ system, $m$,
and the cosine of the spin analyser angle $\theta_H$,
defined as the angle in the $\pi^{\pm}\pi^0$ rest frame
between the \piz\  momentum and the boost
of the $\pi^{\pm}\pi^0$ system. 
The sample is divided into ten equal bins of $\cos\theta_H$ in
the range from $-1$ to $+1$, for
six intervals of \xE\  in the range from 0.025 to 1.
The size of the bins in $x_E$ and $\cos\theta_H$ are large
compared to the experimental resolutions on these quantities,
which are dominated by the $\pi^0$ energy resolution.
The Monte Carlo simulations provide realistic estimates
of these resolutions as they reproduce well the width
of the $\pi^0$ mass peak~\cite{bib-opalpiz}.
They predict that the $\pi^0$ energy resolution varies between
4\% and 8\% over the energy range relevant for the present
analysis, resulting in a resolution of approximately 0.04
on $\cos\theta_H$.

The number of reconstructed \rpm\  mesons in each $\cos\theta_H$
bin and \xE\  interval is obtained from a fit to the
corresponding invariant mass distribution using the
method described in Ref.~\cite{bib-opalpiz}.
These numbers are corrected by the efficiency evaluated by
applying the same reconstruction method and fit procedure to a sample
of 6.4 million hadronic \Zzero\   decays simulated~\cite{bib-gopal}
using the Monte Carlo programs JETSET 7.3 and 7.4~\cite{bib-jetset}
tuned to reproduce the global features of hadronic events as
observed by OPAL~\cite{bib-tuneold,bib-tunenew}.

\begin{figure}[tbp]
\epsfig{file=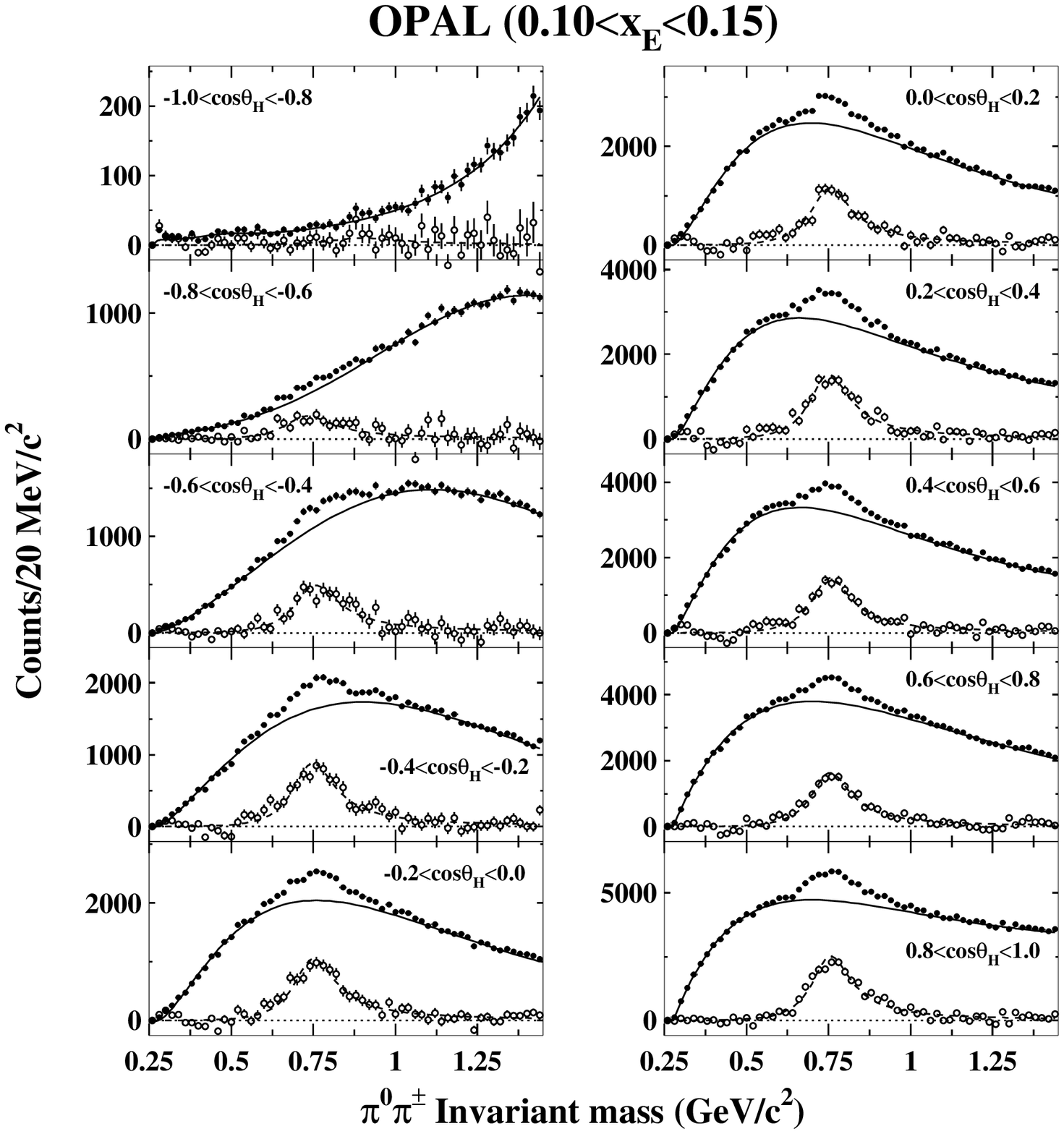
,height=17cm,bbllx=0cm,bblly=0cm,bburx=21cm,bbury=21cm}
\caption  { \label{fig-rhoall}
Invariant mass distributions of $\pi^{\pm}\pi^0$ combinations
for the range 0.10 $<$ $x_E$ $<$ 0.15, for the ten equal bins of
$\cos\theta_H$ between $-1$ and $+1$.
The black circles are the data.
The full lines are the background obtained in the fits to
the data.
The white circles show the \rpm\  signal, scaled by a factor 2,
obtained by subtracting the fitted background (full lines) from
the data.
The dashed lines are the fitted signals scaled by the same factor.
}\end{figure}

\begin{figure}[tbp]
\epsfig{file=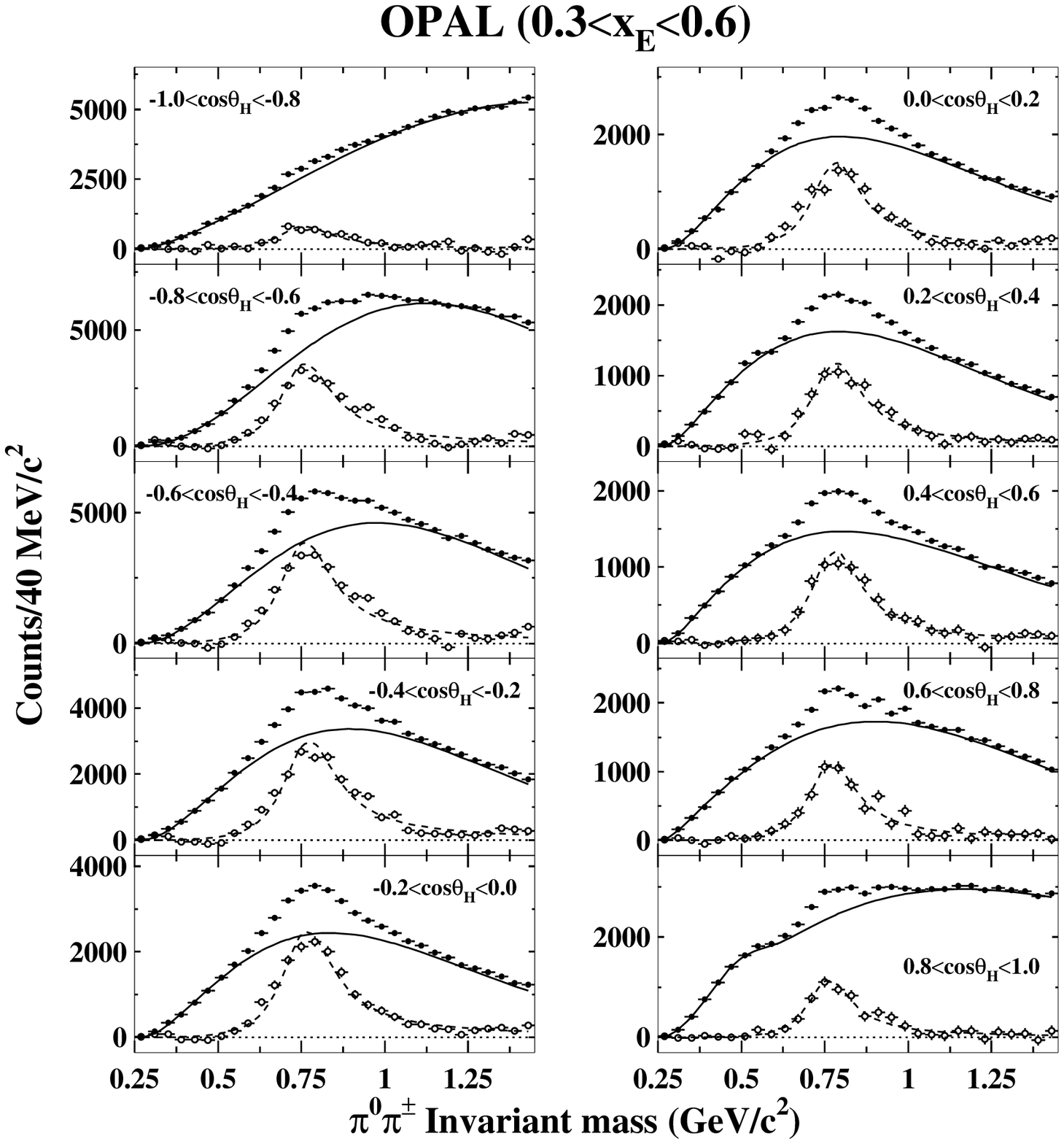
,height=17cm,bbllx=0cm,bblly=0cm,bburx=21cm,bbury=21cm}
\caption  { \label{fig-rhohi}
Invariant mass distributions of $\pi^{\pm}\pi^0$ combinations
for the scaled energy range 0.3 $<$ $x_E$ $<$ 0.6,
for the ten equal bins of $\cos\theta_H$ between $-1$ and $+1$.
The black circles are the data.
The full lines are the background obtained in the fits to
the data.
The white circles show the \rpm\  signal, scaled by a factor 2,
obtained by subtracting the fitted background (full lines) from
the data.
The dashed lines are the fitted signals scaled by the same factor.
}\end{figure}

Examples of invariant mass distributions and fit results
are shown in Figs.~\ref{fig-rhoall} and~\ref{fig-rhohi}.
As can be seen, the shapes of the \rpm\  signal and of the
underlying background vary significantly as a function of
\xE\  and $\cos\theta_H$.
According to the Monte Carlo simulations, this
is due in large part to the dependence of the \piz\ 
reconstruction efficiency on its energy.
To track these effects carefully the fit procedure,
described in Ref.~\cite{bib-opalpiz} and summarised below,
is applied independently to each mass distribution.

In the fit,
the \rpm\  signal is parameterized as a relativistic Breit-Wigner
convoluted with the experimental mass resolution and
multiplied by a factor
$ 1 + C \frac{m_0^2-m^2}{m\Gamma} $
where $\Gamma$ is the width of the resonance~\cite{bib-pdg} and
$m_0$ is the Breit-Wigner pole mass.
This factor has been shown to take adequately into
account shape distortions due to interferences and
residual Bose-Einstein correlations~\cite{bib-lafferty}.
The systematic variations of the signal shape consist of
fixing $C$ to zero or leaving it as a free
parameter and, in addition, the mass
resolution is set to the Monte Carlo prediction or left
as a free parameter.
The systematics associated with the background
are evaluated using two methods.
In the first, the background shape is taken from the simulation and
normalised to the number of counts outside the signal region.
In the second method, the background is parameterized as:
\begin{eqnarray} \label{eqn-rhobkg}
   f(m) & = & p_1 (m-m_{th})^{p_2} \times
       \exp( p_3(m-m_{th}) + p_4 (m-m_{th})^2 )
   \mbox{ , }
\end{eqnarray}
where $m_{th}$ = $m_{\pi^{\pm}}+m_{\pi^0}$
and the parameters $p_1$ to $p_4$ are determined in the fits to the data.
A Gaussian representing the reflection from \omegapipipi\  decays
is added to this shape.
Its width is fixed to the Monte Carlo prediction while its
amplitude and centroid are left as free parameters
in order to absorb possible imperfections in the modelling of the
background near the $\pi^{\pm}\pi^0$
threshold.
The \piz\  selection is also varied as in Ref.~\cite{bib-opalpiz},
testing the sensitivity of the results to
an increase and a decrease of the acceptance by a factor 2.

\subsection{Extraction of the matrix element {\boldmath $\rho_{00}$}
            of {\boldmath $\rho^{\pm}$} mesons}
\label{sect-anarho-extr}

\begin{figure}[tbp]
\epsfig{file=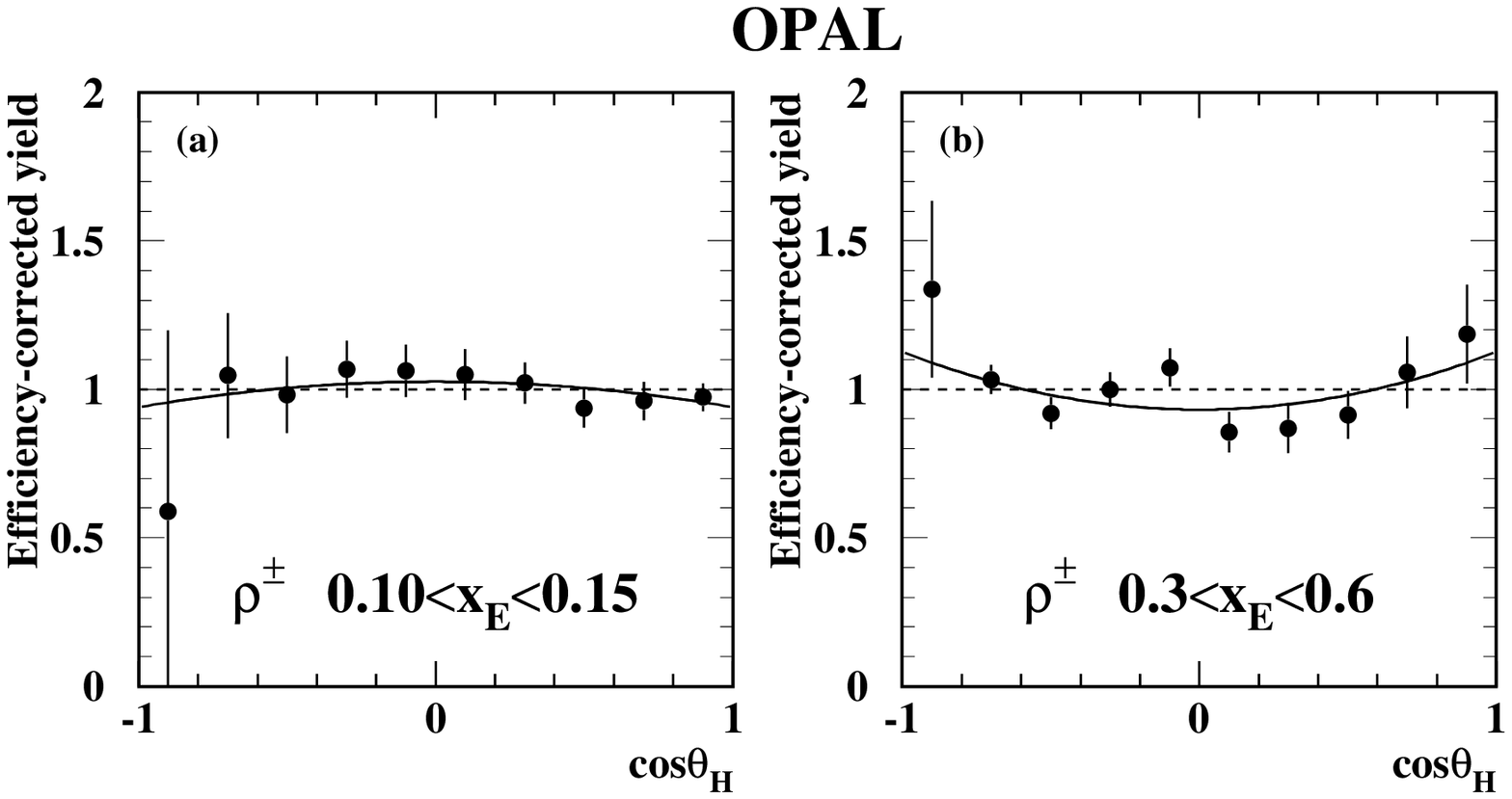
,height=17cm,bbllx=0cm,bblly=0cm,bburx=21cm,bbury=21cm}
\caption  { \label{fig-rhofit}
Efficiency-corrected \rpm\  yields obtained from the fits to the data
shown (a) in Fig.~1 ($0.10<x_E<0.15$) and (b) in Fig.~2 ($0.3<x_E<0.6$).
For clarity, the total yields are normalised to 1 using the results
of the fit to the data in the entire $\cos\theta_H$ range.
The errors are statistical only. The full lines represent the fit
of Eqn.~\protect\ref{eqn-ab} to the data.
}\end{figure}

For each interval of \xE,
the efficiency-corrected  \rpm\  yields are evaluated
for all ten $\cos\theta_H$ bins and fitted to the expression
\begin{eqnarray}
\label{eqn-ab}
 I(\cos\theta_H) & = & A ( 1 + B \cos^2\theta_H ) \; \mbox{ ,}
\end{eqnarray}
From which the value of $\rho_{00}$ is obtained via:
\begin{eqnarray}
\label{eqn-convert}
\rho_{00} & = & \frac{1+B}{3+B}  \; \mbox{ .}
\end{eqnarray}
The parameters $A$ and $B$ and their errors are obtained
from a linear least-squares fit.
As an example, Fig.~\ref{fig-rhofit} shows the fits
to the data of Figs.~\ref{fig-rhoall} and
~\ref{fig-rhohi}.
The fit of Eqn.~\ref{eqn-ab} to the data is repeated for all
systematic variations of the signal and background parameterization,
of the $\pi^0$ selection, and of the Monte Carlo sample used
for the efficiency. The resulting $B$ values are averaged
and the systematic error associated to each source of uncertainty
is taken from the rms deviation from the average.
In this way, these errors reflect the uncertainty on
$B$ and are independent of global variations
which affect only the parameter $A$.

The measured $\rho_{00}$ values are listed in Table~\ref{tab-rho},
together with the statistical errors and the errors from the
following systematic uncertainties:
\begin{enumerate}
\item{The statistical error on the Monte Carlo samples used to
      calculate the efficiency.}
\item{The bias induced by the presence of the background under
      the signal peak. This is estimated by the difference of the
      results of fits to the invariant mass spectra of Monte Carlo
      samples where the background is included or excluded.}
\item{The variations of the fitted yields when the parameterization
      of the shape of the signal and the background is varied as
      in Ref.~\cite{bib-opalpiz}.}
\item{The difference in efficiency obtained with Monte Carlo
      samples using the JETSET tune parameters of
      Refs.~\cite{bib-tuneold} and~\cite{bib-tunenew}.}
\item{The variation observed when the analysis is repeated
      with different values for the cut on the \piz\  selection
      variable~\cite{bib-opalpiz}, corresponding to changes by
      factors from 1/2 to 2 in the acceptance.
      As shown in Ref.~\cite{bib-opalpiz}, the variation of
      this cut induces significant changes in the shape of
      the background and therefore provides an additional
      test of the stability of the results relative to
      the assumptions regarding its shape.}
\end{enumerate}
%

%%%%%%%%%%%%%%%%%%%%%%%%%%%%%%%%%%%%%%%%%%%%%%%%%%%%%%%%%%%%%%%%%%%%%%%%%%%%
                                                           
\section{ Helicity density matrix element 
          {\boldmath $\rho_{00}$} 
          of {\boldmath $\omega$} mesons}
\label{sect-anaom}

The decay \omegapipipi\  has a branching ratio of
88.8$\pm$0.7\%~\cite{bib-pdg}.
In the rest frame of the $\pi^+\pi^-\pi^0$ system,
the momenta of the three pions are in a plane.
The appropriate spin analyser, $\theta_H$, in this
case is the angle between the normal to this plane and
the boost direction~\cite{bib-coseqn},
and Eqn.~\ref{eqn-coseqn}
applies here~\cite{bib-berman} also.

\subsection{{\boldmath $\omega$} meson reconstruction}

The reconstruction of the decay \omegapipipi\
follows exactly that described in Ref.~\cite{bib-opalpiz}.
Charged and neutral pion candidates are selected
as in the \rpm\  analysis.
All triplets comprising two oppositely charged pions and
one neutral pion are considered.
From the energy and momenta of the $\pi^+$, $\pi^-$ and $\pi^0$
candidates three quantities are evaluated:
the scaled energy of the triplet, \xE,
the invariant mass of the \threepi\  system, $m$,
and the cosine of the spin analyser angle, $\theta_H$,
defined above.
The sign of $\cos\theta_H$ is arbitrary and
only its absolute value is considered.
The Monte Carlo simulations predict that the resolution on
$|\cos\theta_H|$ is approximately 0.04,
increasing to 0.06 for $0.3<x_E<0.6$.
The sample is divided into six equal bins of $|\cos\theta_H|$ in
the range from 0 to 1, for
six intervals of \xE\  in the range from 0.025 to 1.

\begin{figure}[tbp]
\epsfig{file=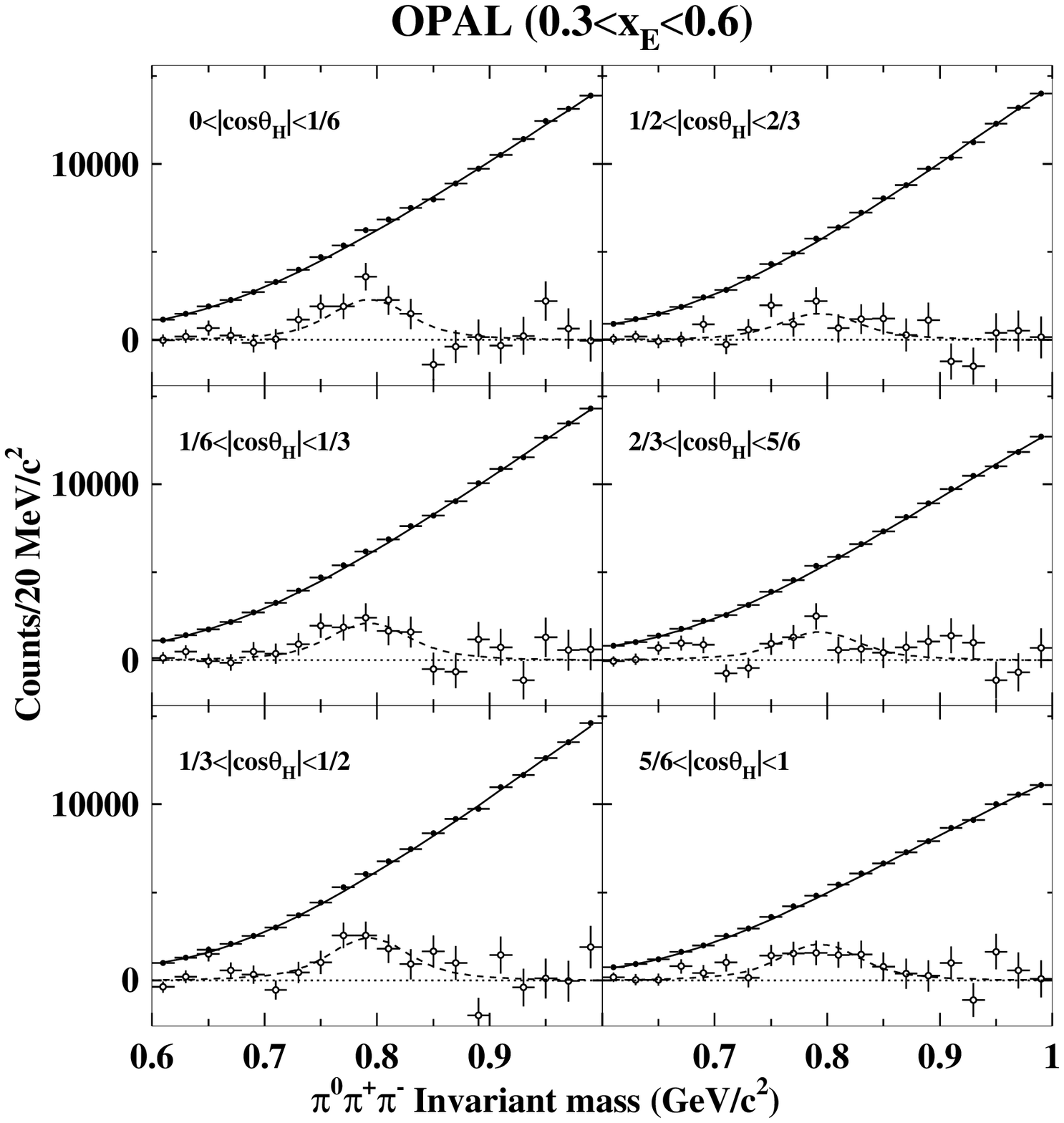
,height=17cm,bbllx=0cm,bblly=0cm,bburx=21cm,bbury=21cm}
\caption  { \label{fig-omehi}
Invariant mass distributions of $\pi^+\pi^-\pi^0$ combinations
for the scaled energy range 0.3 $<$ $x_E$ $<$ 0.6,
for the six equal bins of $|\cos\theta_H|$
between 0 and 1.
The black circles are the data.
The full lines are the background obtained in the fits to
the data.
The white circles show the $\omega$  signal, scaled by a factor 10,
obtained by subtracting the fitted background (full lines) from
the data.
The dashed lines are the fitted signals scaled by the same factor.
}\end{figure}

The number of reconstructed $\omega$ mesons in each $|\cos\theta_H|$
bin and \xE\  interval is obtained from a fit to the
corresponding invariant mass distribution.
Examples of invariant mass distributions and fit results
are shown in Figs.~\ref{fig-omeall} and~\ref{fig-omehi}.
In these fits, the combinatorial background is parameterized
as a third order polynomial, $P_3(m)$.
As in Ref.~\cite{bib-opalpiz}, the $\omega$ signal, $S(m)$,
is described  as the superposition of two
Gaussians sharing the same centroid,
the ratios of the two widths and areas
being determined from the simulation.
According to the simulations, the position and total width
of the peak does not depend on $|\cos\theta_H|$.
Therefore,
for each $x_E$ interval, these two quantities are determined
from a fit to the data in the total interval $0<|\cos\theta_H|<1$,
and are subsequently fixed to these values in the fits to the
six individual $|\cos\theta_H|$ bins.

\subsection{Extraction of the matrix element
            {\boldmath $\rho_{00}$} of {\boldmath $\omega$} mesons}

\begin{figure}[tbp]
\epsfig{file=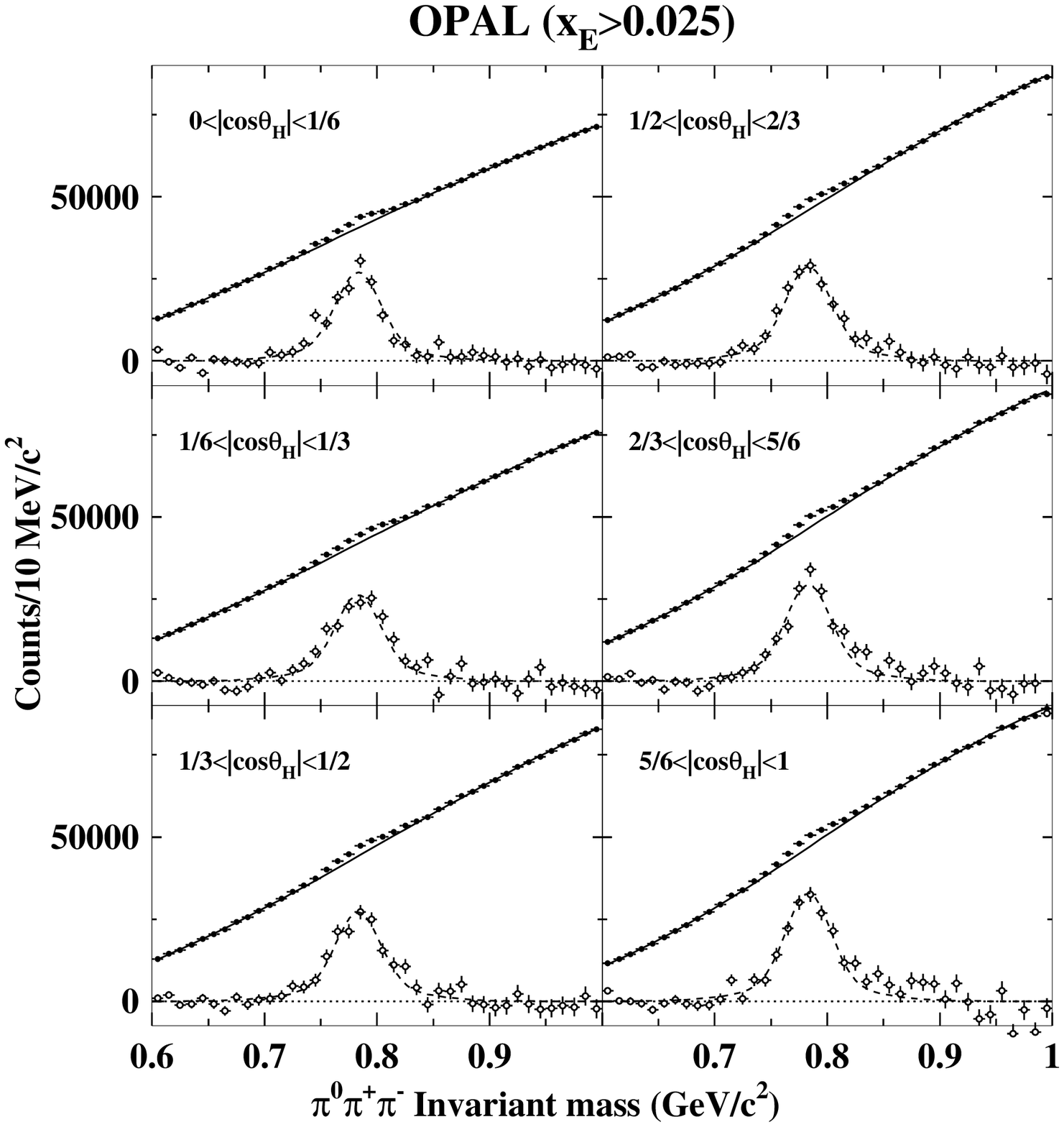
,height=17cm,bbllx=0cm,bblly=0cm,bburx=21cm,bbury=21cm}
\caption  { \label{fig-omeall}
Invariant mass distributions of $\pi^+\pi^-\pi^0$ combinations
for the range $x_E>0.025$, for the six equal bins of $|\cos\theta_H|$
between 0 and 1.
The black circles are the data.
The full lines are the background obtained in the fits to
the data.
The white circles show the $\omega$  signal, scaled by a factor 10,
obtained by subtracting the fitted background (full lines) from
the data.
The dashed lines are the fitted signals scaled by the same factor.
}\end{figure}

\begin{figure}[tbp]
\epsfig{file=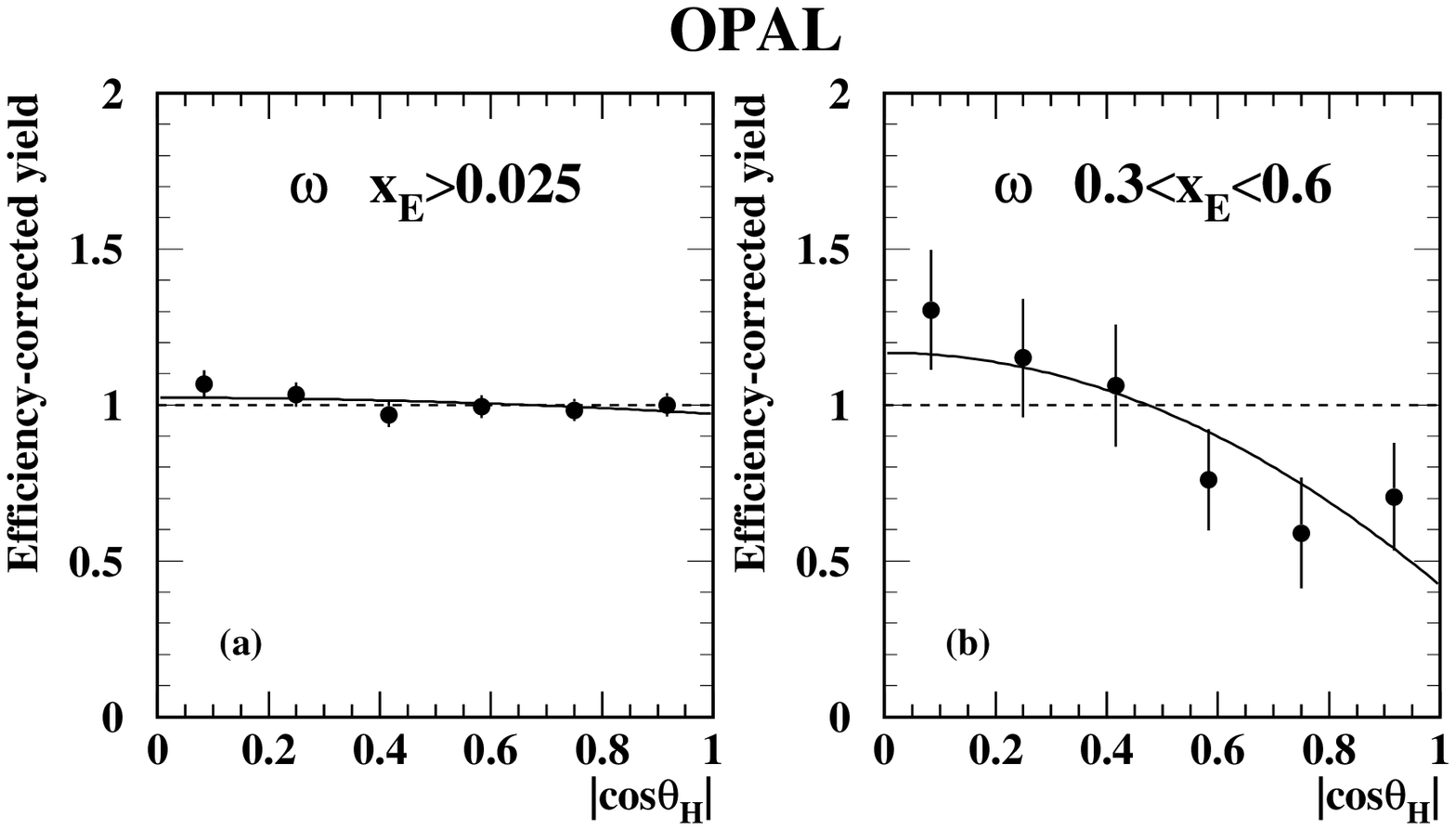
,height=17cm,bbllx=0cm,bblly=0cm,bburx=21cm,bbury=21cm}
\caption  { \label{fig-omefit}
Efficiency-corrected $\omega$ yields obtained from the fits to the data
shown (a) in Fig.~5 ($x_E>0.025$) and (b) in Fig.~6 ($0.3<x_E<0.6$).
For clarity, the total yields are normalised to 1 using the results
of the fit to the data in the entire $| \cos\theta_H |$ range.
The errors are statistical only. The full lines represent the fit
of Eqn.~\protect\ref{eqn-ab} to the data.
}\end{figure}

The extracted $\omega$ yields are corrected by the reconstruction
efficiency evaluated in a similar manner to that in the \rpm\  analysis
(Section~\ref{sect-anarho-extr}).
For each interval in \xE,
the efficiency-corrected $\omega$  yields are evaluated
for the six $|\cos\theta_H|$ bins and fitted to Eqn.~\ref{eqn-ab}.
As an example, Fig.~\ref{fig-omefit} shows the fits
to the data of Figs.~\ref{fig-omeall} and
~\ref{fig-omehi}.
The shape of the background does not vary with $x_E$ as much
as in the case of the \rpm, and here a fit to the sum of all
$x_E$ bins (Fig.~\ref{fig-omeall}) can be performed.
This allows a high-statistics test of the fitting procedure.
The measured $\rho_{00}$ values are listed in Table~\ref{tab-omega},
together with the statistical and systematic uncertainties.
The systematic errors are evaluated as in
Section~\ref{sect-anarho-extr}.
In the case of the $\omega$, the systematic variation of the
shape of the signal (third error in table~\ref{tab-omega})
are derived from determining the width of the $\omega$ peak
from the simulation instead of the data.
As in Ref.~\cite{bib-opalpiz},
an additional error (the sixth in table~\ref{tab-omega})
is derived from the test of whether the extracted $\omega$ signal
has the expected dependence on the matrix element of the
$\omega$ decay~\cite{bib-dkmatrix},
\begin{eqnarray}
\lambda_{\omega} & = &
  \frac{ |\vec{p}^{\,*}_{-}\times\vec{p}^{\,*}_{+}|^2_{\phantom{max}} }
       { |\vec{p}^{\,*}_{-}\times\vec{p}^{\,*}_{+}|^2_{max} }
\end{eqnarray}
where $\vec{p}^{\,*}_{\pm}$ is the momentum of the $\pi^{\pm}$
meson in the \threepi\  rest frame.
The $\lambda_{\omega}$ distribution is expected to rise linearly
with $\lambda_{\omega}$ for the signal,
and be flat for the combinatorial background.
The data are therefore divided into six bins in $\lambda_{\omega}$,
and the two-dimensional distributions $I(m,\lambda_{\omega})$
are fitted with the expression:
\begin{eqnarray}
I(m,\lambda_{\omega}) & = & \frac{P_3(m) + \lambda_{\omega}S(m)}
{ \epsilon(m,\lambda_{\omega}) \; (1+\lambda_{\omega}\delta) }
\end{eqnarray}
where $\epsilon(m,\lambda_{\omega})$ is the acceptance of the
detector determined from the combinatorial background distributions
in the simulations, and $\delta$ is a free parameter in the fit,
designed to account for possible deviations between the
$\lambda_{\omega}$-dependence of the acceptance in the data and in the
Monte Carlo.
The value of the sixth error in Table~\ref{tab-omega} is the
difference between the results obtained from the fits to the
$I(m,\lambda_{\omega})$ and $I(m)$ distributions,
where $I(m)$ is the sum of the six $\lambda_{\omega}$ bins.

%%%%%%%%%%%%%%%%%%%%%%%%%%%%%%%%%%%%%%%%%%%%%%%%%%%%%%%%%%%%%%%%%%%%%%%%%%%%

\section{Results and discussion}
\label{sect-results}

Tables~\ref{tab-rho} and~\ref{tab-omega} present
the measured $\rho_{00}$ values for the \rpm\  and $\omega$
mesons, respectively, together with their statistical,
systematic and total errors.
The total numbers of meson candidates are also listed,
together with their statistical errors.
The statistical errors are dominated by the contribution of
the background under the \rpm\  and $\omega$ signals.
The highest accessible $x_E$ interval is $0.3 < x_E < 0.6$.
The uncertainty on the \rpm\  and $\omega$ samples
with $x_E>0.6$ (approximately 400~$\pm$~200 and
800~$\pm$~400 candidates, respectively) is too large for
a meaningful extraction of $\rho_{00}$.
The systematic errors are evaluated using the method of
Ref.~\cite{bib-opalpiz}.
In Tables~1 and~2,
the scatter of the error values as a function of $x_E$ for some
uncertainty sources indicate that their evaluation may still
be affected by statistical fluctuations.
However the observed scatter is smaller than the size of the
total errors and thus should not affect significantly
the final results.

\begin{figure}[tbp]
\epsfig{file=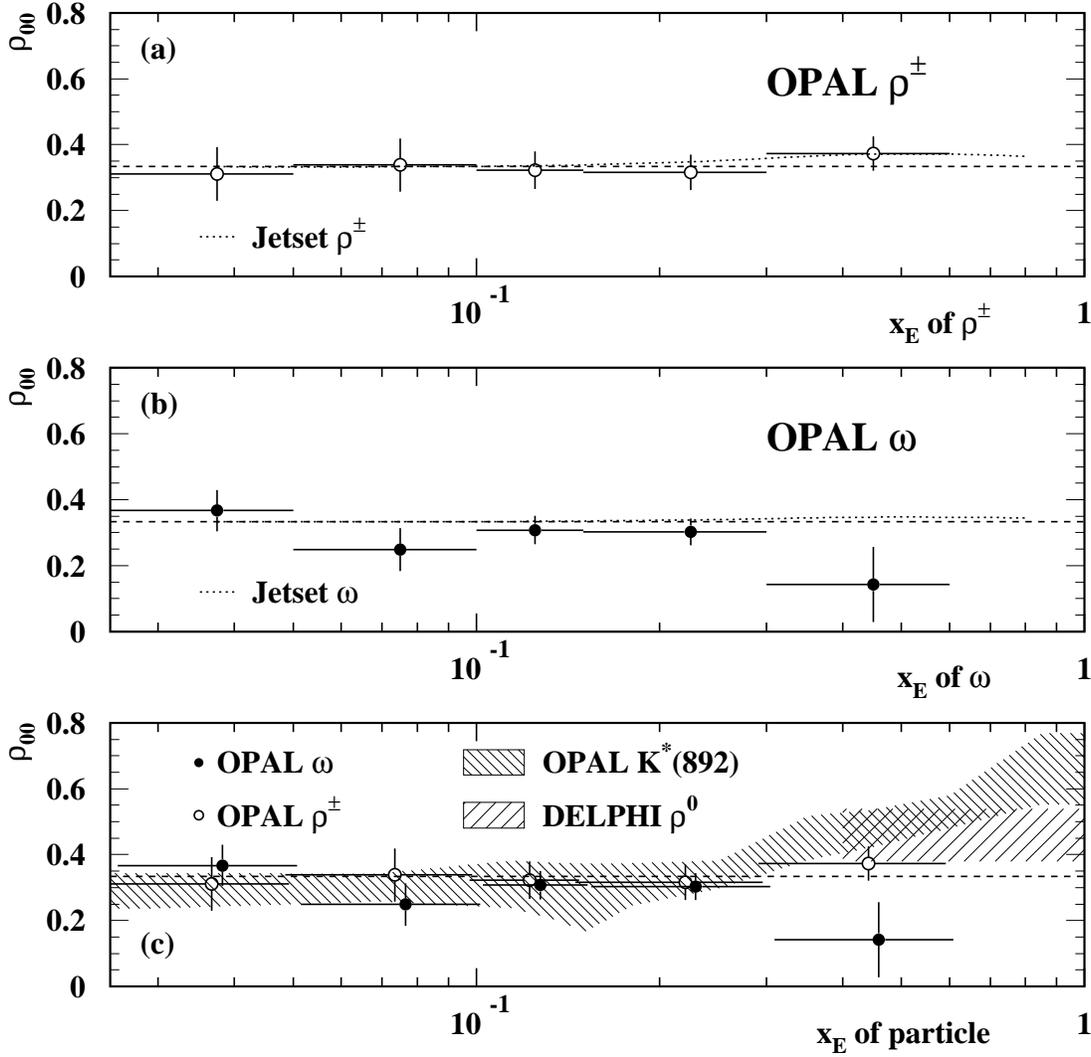
,height=17cm,bbllx=0cm,bblly=0cm,bburx=21cm,bbury=21cm}
\caption  { \label{fig-results}
Measured $\rho_{00}$ values as a function of $x_E$ for (a)
\rpm\  mesons and (b) $\omega$ mesons produced in Z$^0$ decay.
The dashed lines correspond to an isotropic spin distribution
(1/3).
The dotted lines represent
the amount of alignment expected from the $J^P=0^-\rightarrow0^-+1^-$
decays present in the simulations described in Sect.~4.
%the calculated $\rho_{00}$ values
%assuming that $J^P=0^-\rightarrow0^-+1^-$ decays are the only
%source of alignment, and that the number of these decays per
%hadronic Z$^0$ decays is as predicted by the
%Monte Carlo simulations.
In (c), the data from (a) and (b) are compared with other
measurements~\protect\cite{bib-alidelphi,bib-aliopks}.
}\end{figure}

The measured values of $\rho_{00}$ as a function of $x_E$
are shown in Figs.~\ref{fig-results}a and b for the \rpm\
and $\omega$ mesons, respectively.
The measurements are compatible with 1/3
(the dashed line in Figs~\ref{fig-results}a and~b),
corresponding to
a statistical mix of helicity $-1$, 0 and $+1$ states.
Out of ten measurements, eight are equal to 1/3 within
one standard deviation, and two are within two
standard deviations.
In other cases where vector meson spin alignment has been
observed~\cite{bib-aliopal,bib-alidelphi,bib-aliopks}
it appears only at meson energies above $x_E>0.3$.
Here, in the energy range 0.3 $<$ $x_E$ $<$ 0.6,
the values of $\rho_{00}$ for the \rpm\  and $\omega$
mesons are \rhoreslhi\  and \omreslhi.

According to the Monte Carlo, and depending on the \rpm\  energy,
up to 10\% of \rpm\  originate from the sequential decays of
$J^P$ = $0^-$ mesons into a \rpm\  ($J^P$ = $1^-$) and
another $J^P$ = $0^-$ meson.
The most important source of these decays is
D$^0$ $\rightarrow$ $\rho^+$  K$^-$.
In these decays the orbital angular momentum of the \rpm\  
must be opposite to its intrinsic spin so that
the \rpm\  must be in a helicity 0 state
($\rho_{00}$ = 1) in the rest frame of the parent meson.
The degree of alignment in the laboratory frame
depends on the relative momenta of the \rpm\
and the parent 0$^-$ meson.
Simulations predict that only 60\% of the alignment
survives in the laboratory frame for \rpm\  mesons
in the interval $0.3<x_E<0.6$ and that it
essentially disappears for $x_E<0.1$.
The effect of the alignment of the \rpm\  mesons
coming from $J^P=0^-\rightarrow0^-+1^-$ decays
present in the JETSET 7.3 and 7.4 samples based on the
parameters of Ref.~\cite{bib-tuneold} and~\cite{bib-tunenew}
are almost indistinguishable, despite the inclusion
of $L=1$ mesons in the latter simulation.
This prediction is shown in Fig.~\ref{fig-results}a
as a dotted line.
It agrees with the data.
However, these small deviations from 1/3 at high $x_E$ are
comparable in size to the experimental errors.
For the $\omega$, the simulations predict that less
than 5\% of $\omega$ mesons in the interval $0.3<x_E<0.6$
come from $J^P=0^-\rightarrow0^-+1^-$ decays.
The expected impact on the average $\omega$ alignment,
shown as a dotted line in Fig.~\ref{fig-results}b,
is significantly smaller than in the \rpm\  case
(Fig.~\ref{fig-results}a).
If the contributions of the expected sources of
$J^P=0^-\rightarrow0^-+1^-$ decays
are removed, the remaining population is
still compatible with $\rho_{00}$ = 1/3 over the entire $x_E$ range.

Fig.~\ref{fig-results}c compares the \rpm\  and $\omega$
measurements with those for the K$^*(892)^0$ obtained
by OPAL~\cite{bib-aliopks}
and the $\rho^0$ by DELPHI~\cite{bib-alidelphi}.
The less precise DELPHI K$^*(892)^0$
data~\cite{bib-alidelphi} are not shown for
clarity: they are consistent with the OPAL data.
The K$^*(892)^0$ data show a significant preference for
$\rho_{00}$ values above 1/3 at values of $x_E$ above 0.3.
By itself, the $\rho^0$ data was not sufficiently precise to conclude
whether similar alignment values were also observed for
light, unflavoured
mesons in the corresponding $x_E$ range.

\begin{figure}[tbp]
\epsfig{file=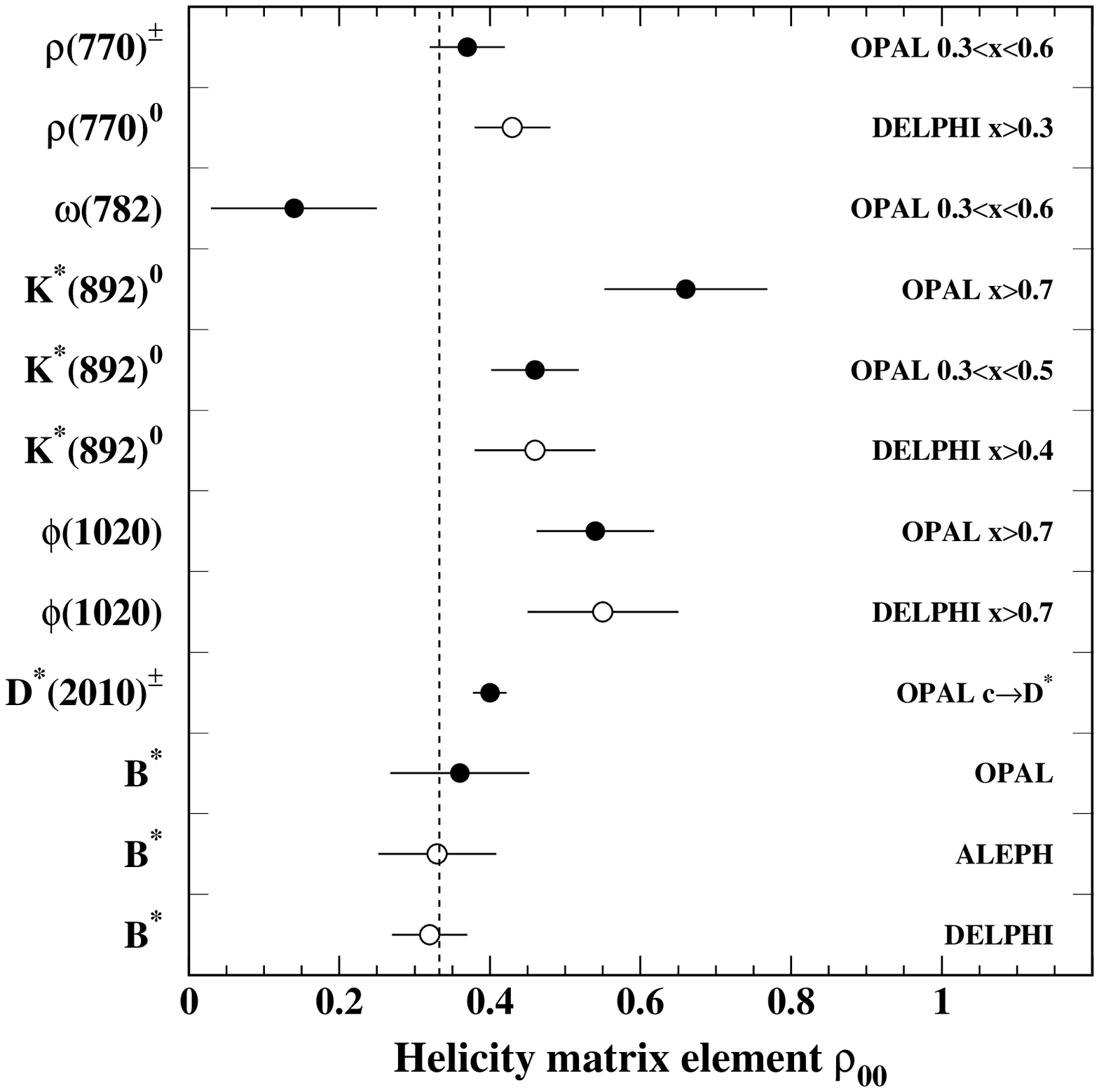
,height=17cm,bbllx=0cm,bblly=0cm,bburx=21cm,bbury=21cm}
\caption  { \label{fig-compare}
Summary of published $\rho_{00}$ measurements for vector
mesons produced in Z$^0$ decays.
The results for the \rpm\  and $\omega$ are from this work.
The Delphi results for the $\rho^0$, K$^*(892)^0$ and $\phi$
mesons are from Ref.~\cite{bib-alidelphi}, and the OPAL results for
the  $\phi$, D$^{*\pm}$ and B$^*$ mesons are from
 Ref.~\cite{bib-aliopal}.
Other results are from Refs.~\cite{bib-aliopks} (OPAL K$^*(892)^0$)
and~\cite{bib-alibstar} (ALEPH and DELPHI B$^*$).
Only measurements at large $x$ or corresponding to primary quarks
are shown.
The K$^*(892)^0$ data at $0.3<x_E<0.5$ and $x_E>0.7$ are also
shown as an example of the possible $x_E$ dependence.
}\end{figure}

Fig.~\ref{fig-compare} shows a compilation of $\rho_{00}$ measurements
for different mesons in different $x_E$ ranges.
Up to now, the B$^*$ meson was the only case where $\rho_{00}$=1/3
was clearly preferred.
The K$^*(892)^0$ and
$\phi$ mesons\footnote{The OPAL $\phi$ data~\cite{bib-aliopks}
are corrected for the effects of
$J^P=0^-\rightarrow0^-+1^-$ decays
predicted by the Monte Carlo simulations.
However the size of the correction is small compared
to the observed alignment.}
appeared to prefer larger values
of $\rho_{00}>1/2$ at high $x_E$, with the D$^*$ mesons in between.
In contrast,
the new results on the \rpm\  and $\omega$,
together with the previous DELPHI $\rho^0$ results,
appear to prefer $\rho_{00}$ values close to 1/3.
Therefore the presence of spin alignment above $x_E>0.3$
cannot be considered to be
a general property of mesons produced in hadronic Z$^0$ decays.
This could either be due to the influence of cascade decays on
the observed alignments or to some unknown mechanism producing
the alignment.
More measurements above $x_E=0.3$,
and particularly above 0.6,  would contribute significantly to the
understanding of meson spin alignment.

%%%%%%%%%%%%%%%%%%%%%%%%%%%%%%%%%%%%%%%%%%%%%%%%%%%%%%%%%%%%%%%%%%%%%%%%%%%%

\section{Conclusion}
\label{sect-conclusion}

The helicity density matrix elements $\rho_{00}$ of \rpm\
and $\omega$ mesons
produced in Z$^0$ decays have been measured using the
OPAL detector at LEP.
Over the entire energy range, 
the measured values are compatible with 1/3
corresponding to
a statistical mix of helicity $-1$, 0 and $+1$ states.
The measurements in the highest accessible
energy range 0.3 $<$ $x_E$ $<$ 0.6
are \rhoreshi\  and  \omreshi\
for \rpm\   and $\omega$ mesons, respectively.
Taken together, these results are lower than the values observed
at high $x_E$ for the K$^*(892)^0$ and $\phi$ mesons.
 
%%%%%%%%%%%%%%%%%%%%%%%%%%%%%%%%%%%%%%%%%%%%%%%%%%%%%%%%%%%%%%%%%%%%%%%%%%%%
\newpage
\par
{\Large {\bf Acknowledgements:}}
\par
We particularly wish to thank the SL Division for the efficient operation
of the LEP accelerator at all energies
 and for their continuing close cooperation with
our experimental group.  We thank our colleagues from CEA, DAPNIA/SPP,
CE-Saclay for their efforts over the years on the time-of-flight and trigger
systems which we continue to use.  In addition to the support staff at our own
institutions we are pleased to acknowledge the  \\
Department of Energy, USA, \\
National Science Foundation, USA, \\
Particle Physics and Astronomy Research Council, UK, \\
Natural Sciences and Engineering Research Council, Canada, \\
Israel Science Foundation, administered by the Israel
Academy of Science and Humanities, \\
Minerva Gesellschaft, \\
Benoziyo Center for High Energy Physics,\\
Japanese Ministry of Education, Science and Culture (the
Monbusho) and a grant under the Monbusho International
Science Research Program,\\
Japanese Society for the Promotion of Science (JSPS),\\
German Israeli Bi-national Science Foundation (GIF), \\
Bundesministerium f\"ur Bildung, Wissenschaft,
Forschung und Technologie, Germany, \\
National Research Council of Canada, \\
Research Corporation, USA,\\
%Hungarian Foundation for Scientific Research, OTKA T-016660, 
%T023793 and OTKA F-023259.\\
Hungarian Foundation for Scientific Research, OTKA T-029328, 
T023793 and OTKA F-023259.\\

%%%%%%%%%%%%%%%%%%%%%%%%%%%%%%%%%%%%%%%%%%%%%%%%%%%%%%%

%======================================================================== 

\newpage

%======================================================================== 

\begin{table}[p]
\hspace*{6cm}
\begin{sideways}
\begin{minipage}{1.15\textwidth}
\centerline\mbox{
%\begin{table}[tbp]
%\begin{center}
\begin{tabular}{|c||c||c||c|c|c|c|c|c||c|}
\hline
\hline
               & Number
               &        &       & 1    &  2   &   3    &   4   &  5   &     \\
      $x_E$    & of candidates
               & $\rho_{00}$ 
                        &  Stat. 
                                & Stat. 
                                        & Bkg. & Signal & Diff.
                                                                & $\pi^0$
                                                                        & Total \\
      range    & ($\times 10^3$)
               & value
                        &  (data)
                                & (MC)
                                        & bias  & shape & MC 
                                                                & select.
                                                                        & error \\
\hline
% 0.025 - 1.000 & 783 $\pm$ 9 &
%                  0.245 & 0.035 & 0.012 & 0.073 & 0.022 & 0.010 & 0.046 & 0.097 \\
 0.025 - 0.050 &   48 $\pm$ 5 &
                  0.312 & 0.066 & 0.004 & 0.000 & 0.000 & 0.005 & 0.047 & 0.081 \\
 0.050 - 0.100 &  156 $\pm$ 5 &
                  0.338 & 0.060 & 0.030 & 0.030 & 0.008 & 0.004 & 0.033 & 0.081 \\
 0.100 - 0.150 &  222 $\pm$ 4 &
                  0.322 & 0.044 & 0.017 & 0.023 & 0.014 & 0.008 & 0.018 & 0.058 \\
 0.150 - 0.300 &  273 $\pm$ 3 &
                  0.316 & 0.027 & 0.019 & 0.032 & 0.018 & 0.019 & 0.009 & 0.054 \\
 0.300 - 0.600 &   60 $\pm$ 1 &
                  0.373 & 0.035 & 0.029 & 0.001 & 0.011 & 0.009 & 0.022 & 0.052 \\
% 0.600 - 1.000 & 0.4 $\pm$ 0.2 &
%                 -0.291 & 0.381 & 0.161 & 0.336 & 0.061 & 0.130 & 0.234 & 0.600 \\
\hline
\hline
\end{tabular}
\caption{\label{tab-rho}
Measured $\rho_{00}$ values for \rpm\  mesons as a function of $x_E$,
together with the statistical, systematic and total errors.
The different systematic error contributions 1--5 are described in
Sect. 2.2.
The total numbers of $\rho^{\pm}$ candidates in each $x_E$ bin are
also listed, together with their statistical errors, for the $\pi^0$ cut yielding
the largest acceptance. These numbers are only shown to give an idea of the statistical
precision of the measurement and are not used in the analysis.}
%\end{center}
%\end{table}
}
\end{minipage}
\end{sideways}
\end{table}

%======================================================================== 

%\begin{table}[btp]
%\begin{center}
\begin{table}[p]
\hspace*{6cm}
\begin{sideways}
\begin{minipage}{1.15\textwidth}
\centerline\mbox{
\begin{tabular}{|c||c||c||c|c|c|c|c|c|c||c|}
\hline
\hline
               & Number 
               &        &       & 1    &  2   &   3    &   4   &  5   &   6     & \\
      $x_E$    & of candidates
               & $\rho_{00}$ 
                        &  Stat. 
                                & Stat. 
                                        & Bkg.  & Mass  & Diff. & $\pi^0$
                                                                       &        & Total \\
      range    & ($\times 10^3$)
               & value
                        &  (data)
                                & (MC)
                                        & bias  & resol. & MC   & select.
                                                                        & $\lambda_{\omega}$
                                                                                & error \\
\hline
 0.025 - 1.000 & 226.0 $\pm$ 2.7 &
                  0.312 & 0.011 & 0.001 & 0.010 & 0.008 & 0.005 & 0.027 & 0.000 & 0.032 \\
 0.025 - 0.050 &  32.8 $\pm$ 1.2 &
                  0.367 & 0.040 & 0.004 & 0.038 & 0.002 & 0.013 & 0.027 & 0.001 & 0.063 \\
 0.050 - 0.100 & 100.2 $\pm$ 1.7 &
                  0.249 & 0.018 & 0.003 & 0.047 & 0.008 & 0.008 & 0.038 & 0.007 & 0.065 \\
 0.100 - 0.150 &  47.5 $\pm$ 1.2 &
                  0.308 & 0.021 & 0.003 & 0.015 & 0.003 & 0.010 & 0.032 & 0.005 & 0.043 \\
 0.150 - 0.300 &  41.4 $\pm$ 1.0 &
                  0.303 & 0.026 & 0.004 & 0.011 & 0.007 & 0.012 & 0.023 & 0.011 & 0.041 \\
 0.300 - 0.600 &   4.8 $\pm$ 0.4 &
                  0.142 & 0.081 & 0.010 & 0.034 & 0.014 & 0.029 & 0.049 & 0.041 & 0.114 \\
% 0.600 - 1.000 &   0.8 $\pm$ 0.4 &
%                 -0.311 & 0.568 & 0.263 & 0.697 & 0.001 & 0.370 & 0.115 & 0.204 & 1.034 \\
\hline
\hline
\end{tabular}
\caption{\label{tab-omega}
Measured $\rho_{00}$ values for $\omega$  mesons as a function of $x_E$,
together with the statistical, systematic and total errors.
The different systematic error contributions 1--6 are described in
Sect. 2.2 and 3.2.
The total numbers of $\omega$ candidates in each $x_E$ bin are
also listed, together with their statistical errors, for the $\pi^0$ cut yielding
the largest acceptance. These numbers are only shown to give an idea of the statistical
precision of the measurement and are not used in the analysis.}
%\end{center}
%\end{table}
}
\end{minipage}
\end{sideways}
\end{table}

%=============================================================== Figure 1

%=============================================================== Figure 2

%=============================================================== Figure 3

%=============================================================== Figure 4

%=============================================================== Figure 5

%=============================================================== Figure 6

%=============================================================== Figure 7

%=============================================================== Figure 8

%======================================================================== 


\newpage
\begin{thebibliography}{99}

\bibitem{bib-aliopal}
  OPAL Collaboration, K. Ackerstaff {\it et al.}, 
%  {\em Study of $\phi(1020)$, D$^{\pm}$ and B$^*$ Spin alignment
%       in hadronic Z$^0$ decays}, CERN-PPE/97-05
  Z. Phys {\bf C 74} (1997) 437.

\bibitem{bib-alidelphi}
  DELPHI Collaboration, P. Abreu {\it et al.}, 
%  {\em Measurement of the Spin Density Matrix for the $\rho^0$,
%       K$^{*0}(892)$ and $\phi$ Produced in Z$^0$ decays}, CERN-PPE/97-55,
  Phys. Lett. {\bf B 406} (1997) 271. 

\bibitem{bib-aliopks}
  OPAL Collaboration, K. Ackerstaff {\it et al.}, 
%  {\em Spin alignment of leading K$^*(892)^0$ mesons
%       in hadronic Z$^0$ decays},  CERN-PPE/97-094
  Phys. Lett. {\bf B 412} (1997) 210.

\bibitem{bib-alibstar}
  DELPHI Collaboration, P. Abreu {\it et al.}, 
  Z. Phys. {\bf C 68} (1995) 353; \\
  ALEPH Collaboration, D. Buskulic {\it et al.}, 
  Z. Phys. {\bf C 69} (1995) 393.

\bibitem{bib-theo}
  J.F. Donoghue, Phys. Rev. {\bf D 19} (1979) 2806;\\
  J.E. Augustin and F.M. Renard, Nucl. Phys. {\bf B 162} (1980) 341;\\
  A.F. Falk and M.E. Peskin, Phys. Rev. {\bf D 49} (1994) 3320; \\
  M.A. Anselmino, M. Bertini, F. Murgia and P. Quintairos,
  Eur. Phys. J. C {\bf 2} (1998) 539;\\
  and references therein.

\bibitem{bib-jetset}
   B. Andersson, G. Gustafson, G. Ingelman and T. Sj\"{o}strand,
   Phys. Rep. {\bf 97} (1983) 31; \\
   T. Sj\"{o}strand, Comp. Phys. Comm. {\bf 39}
   (1986) 347; \newline
   T. Sj\"{o}strand and M. Bengtsson, Comp. Phys. Comm. {\bf 43}
   (1987) 367; \newline
   T. Sj\"{o}strand, Comp. Phys. Comm. {\bf 82} (1994) 74.

\bibitem{bib-herwig}
   B.R. Webber, Nucl. Phys. {\bf B 238} (1984) 492.;\\
   G. Marchesini and B.R. Webber, Nucl. Phys.
   {\bf B 310} (1988) 461; \\
   G. Marchesini {\it et al.}, Comp. Phys. Comm.
   {\bf 67} (1992) 465.

\bibitem{bib-opaldet}
   OPAL Collaboration, K. Ahmet {\it et al.}, Nucl. Instr. and Meth.
   {\bf A 305} (1991) 275.

\bibitem{bib-opalsi}
   P.P.~Allport {\it et al.}, Nucl. Instr. and Meth.
   {\bf A 324} (1993) 34; \\
   P.P.~Allport {\it et al.}, Nucl. Instr. and Meth.
   {\bf A 346} (1994) 476.

\bibitem{bib-dedx}
   M. Hauschild {\it et al.}, Nucl. Instr. and Meth.
   {\bf A 314} (1992) 74.

\bibitem{bib-opalline}
   OPAL Collaboration, G. Alexander {\it et al.},
   Z. Phys. {\bf C 52} (1991) 175.

\bibitem{bib-opalpiz}
   OPAL Collaboration, K. Ackerstaff {\it et al.}, 
% {\em Photon and Light Mesons Production in Z$^0$ Decays},
   Eur. Phys. J. C {\bf 5} (1998) 411.

\bibitem{bib-coseqn}
   C. Bourelly, E. Leader and J. Soffer, Phys. Rep. {\bf 59}
   (1980) 95.

\bibitem{bib-gopal}
   J. Allison {\it et al.}, Nucl. Instr. and Meth.
   {\bf A 317} (1992) 47.

\bibitem{bib-tuneold}
   OPAL Collaboration, M.Z. Akrawy {\it et al.},
   Z. Phys. {\bf C 47} (1990) 505.

\bibitem{bib-tunenew}
   OPAL Collaboration, G. Alexander {\it et al.},
   Z. Phys. {\bf C 69} (1996) 543.

%\bibitem{bib-opalphi}
%% Inclusive Strange Vector and Tensor Meson Production in Hadronic Z Decays 
%   OPAL Collaboration, R. Akers {\it et al.},
%   Z. Phys. {\bf C 68} (1995) 1.

%\bibitem{bib-delrho}
%% Production Characteristics of K$^0$ and Light Meson Resonances
%% in Hadronic Decays of the Z}
%   DELPHI Collaboration, P. Abreu {\it et al.},
%   Z. Phys. {\bf C 65} (1995) 587.


\bibitem{bib-pdg}
  The Particle Data Group, C. Caso {\em et al.},
  Eur. Phys. J. C {\bf 3} (1998) 1.


\bibitem{bib-lafferty}
% Residual Bose-Einstein correlations in inclusive $\pi+\pi-$ systems and
% the $\rho(770)^0$ lineshape in multihadronic Z$^0$ decays
  G.D. Lafferty, Z. Phys. {\bf C 60} (1993) 659.

%\bibitem{bib-omhel}
%  Atkison {\it et al.}, Nucl. Phys. {\bf B 231} (1984) 15.
\bibitem{bib-berman}
  S.M. Berman and M. Jacob, Phys. Rev. {\bf 139} (1965) B 1023.

\bibitem{bib-dkmatrix}
% Spin and Parity of the $\omega$ Meson
   M.L. Stevenson {\it et al.}, Phys. Rev. {\bf 125} (1962) 687.

%\bibitem{bib-opalrho}
%   OPAL Collaboration, P.D. Acton {\it et al},
%   Z. Phys. {\bf C 56} (1992) 521.


\end{thebibliography}
\end{document}